\documentclass[aps,prb,twocolumn,superscriptaddress,showpacs]{revtex4}
\usepackage{graphicx}
\usepackage{dcolumn}% Align table columns on decimal point
\usepackage{color}
\usepackage[breaklinks,colorlinks = true,linkcolor = blue,
urlcolor=blue,citecolor=blue]{hyperref}

\begin{document}
\title{Spin-orbit and impurity scattering in an integrable electron model: \\ 
Exact results for dynamic correlations}

\author{A.A. Zvyagin}
\affiliation{Max-Planck Institut f\"ur Physik komplexer Systeme, Noethnitzer
Str., 38, D-01187, Dresden, Germany}
\affiliation{B.I.~Verkin Institute for Low Temperature Physics and Engineering 
of the National Academy of Sciences of Ukraine, Lenin Ave., 47, Kharkov, 
61103, Ukraine}
\author{H. Johannesson}
\affiliation{Department of Physics, University of Gothenburg, SE-412 96 
Gothenburg, Sweden}

\begin{abstract}
We introduce an integrable model of spin-polarized interacting 
electrons subject to a spin-conserving spin-orbit interaction. Using the Bethe ansatz and conformal field theory we calculate the exact large-time single-electron and density correlations and find that while the spin-orbit interaction enhances the single-electron Green's function, the density correlations get suppressed. Adding a localized impurity and coupling it to the electrons so that integrability is preserved,  the dynamic correlations are found to change significantly after a quantum quench with the impurity interaction switched on suddenly. When the electrons are confined to a periodic structure, the correlations are indifferent to the location of the impurity and only carry an imprint of its intrinsic properties. We conjecture that this unusual feature originates from the integrability of the model.
\end{abstract}

%\date{\today}
\maketitle

\section{Introduction}
In recent years, there has been a growing interest in materials and solid-state
devices with strong spin-orbit interactions. Being a relativistic effect, a 
spin-orbit  interaction (SOI) reveals itself as a velocity-dependent magnetic 
field acting on the spin of a particle moving in an electric field. This 
enables the polarization and manipulation of carrier spins by electric fields 
only $-$ bypassing design complexities connected with local magnetic fields 
$-$ and is at the heart of current efforts to fuse spintronics with 
semiconductor technologies \cite{AwschalomFlatte}. Spin polarization can be 
generated by an SOI in a variety of ways: impurity scattering (as in 
the anomalous \cite{Nagaosa} and spin Hall effects \cite{DyakonovPerel}), via 
an external electric bias (``current-induced spin polarization'' 
\cite{IvchenkoPikus}), or, topologically, through spin-momentum
locking from strong atomic SOIs  (as in topological insulators 
\cite{KaneZhang,MooreBalents}). Once a 
spin-polarized current is produced, it may then be manipulated by exploiting 
the presence of other SOIs due to broken symmetries from interfaces, crystal 
structures, strain, or electric fields. The generic examples in 
semiconductor heterostructures are the Rashba and Dresselhaus SOIs 
\cite{D}.

In many proposals for spintronic devices, the interaction between electrons has 
to be taken into account, hence, it is important to investigate the effects of SOIs 
together with electron-electron interactions. This is particularly so
for low-dimensional structures where fluctuations are strongly enhanced due to 
nonanalyticities in the density of states. The additional presence of 
impurities and disorder leads to a complex problem, making non-perturbative 
theoretical results highly desirable. 

In the present work, we make a first attempt at this task by studying an 
exactly solvable model of one-dimensional (1D) interacting electrons subject to 
spin-orbit {\em and} impurity scattering. To allow for an exact solution, we 
study a {\em minimal model} where the electrons are spin-polarized, and with 
the added SOI preserving the spin polarization. To simplify further, we 
consider a single impurity, and we devise its interaction with the itinerant 
electrons in such a way as to make the model integrable, amenable to a Bethe 
ansatz approach. While the resulting interaction becomes rather unwieldy $-$ as 
expected from past work on integrable impurities \cite{IntImp} $-$ it could 
nowadays conceivably be synthesized in a cold atomic gas confined to an optical 
nanotube \cite{Catani}. Indeed, the study of synthetic SOIs in cold atomic 
gases, mimicking effects from semiconductor physics, is now coming of age 
\cite{GalitskiSpielman}, making this line of research quite timely.

The Bethe ansatz solvability of the model allows us to extract its finite-size spectrum, from which the scaling exponents for correlation functions can be obtained via conformal field theory \cite{Cardy}. Focussing on the large-time dynamical correlations, we find that while the spin-orbit interaction enhances the single-electron Green's function, the density correlations get suppressed. As expected, the presence of the integrable impurity does not influence the scaling exponents at equilibrium: Integrability implies that the impurity supports forward scattering only, with the sole effect that a scattered electron picks up a phase shift that can be absorbed in a twisted boundary condition on its  wave function. Considering a local quantum quench $-$ with the impurity-electron interaction suddenly switched on $-$ one might anticipate that the large-time asymptotics maps onto the equilibrium impurity model and therefore is also insensitive to the presence of the impurity. However, this is not the case. When the electrons are confined to a ring, the scaling exponents for the large-time dynamic bulk correlation exponents {\em do} acquire a dependence on the impurity.  Moreover, the quantum quench tends to boost electron density correlations, whereas the spin-orbit interactions does the opposite. This suggests that the very feature of integrability endows the ground state with a highly quantum entangled structure where also ``far-away'' electrons feel the presence of the impurity. We conjecture that this feature reflects the way in which an integrable impurity embedded in a one-dimensional ring scatters electrons: All electrons are perfectly transmitted across the impurity site, with the quantum quench releasing a finite-momentum excitation that runs around the ring and influences correlations uniformly in the bulk at large times. This salient feature may enable the engineering of quantum states in one-dimensional structures with ``functional quantum impurities" which do not corrupt electron transport and, moreover, could be used to promote electron correlations. \\ 

\section{Model}
We consider a 1D spin-polarized interacting electron system, with the SOI 
coming from an electric field perpendicular to both the spin polarization and 
the direction of electron propagation. Awaiting future cold atom realizations 
\cite{GalitskiSpielman}, such a setup may be materialized using a quantum wire 
patterned in a zinc-blende semiconductor quantum well where shear strain 
gradients emulate an internal electric field \cite{BernevigZhang}, and with the 
device put on top of a ferromagnetic insulator to provide for the spin 
polarization. We should stress, however, that we do not aspire to model a 
particular experiment. Instead, the main reason for the design of 
our model is to obtain a sufficiently simple but nontrivial theory that allows 
for an exact solution. Thus, we take as a Hamiltonian 
${\cal H}_{\text{wire}} = \sum_j h_{j,j+1}$, where 
\begin{equation}
h_{j,j+1} = (t+i\alpha)c^{\dagger}_jc_{j+1} +{\rm h.c.} + V n_jn_{j+1} - \mu n_j 
\ , 
\label{Hhom}
\end{equation}
where $c^{\dagger}_j$ ($c_j$) is the creation (destruction) operator for an 
electron at the $j$th site, $n_j = c^{\dagger}_jc_j$,  $t$ is the hopping 
amplitude in the absence of an SOI, $\alpha$ is the SOI amplitude, and $V$ 
is the interaction strength between electrons at neighboring sites (see Fig. 1). The 
hopping term in Eq. (\ref{Hhom}) can be re-written \cite{MGE} as 
$t'(e^{i2\pi \phi} c^{\dagger}_jc_{j+1} +{\rm h.c.})$, where 
$t'=\sqrt{t^2+\alpha^2}$, and $\tan (2\pi \phi) = \alpha/t$, and one then
recognizes ${\cal H}_{\text{wire}}$ as a 1D analog of the Haldane-Hubbard 
model \cite{HH}. With the help of a gauge transformation the phase factor can 
be removed completely from the theory for the case of an open chain, and 
transferred to twisted boundary conditions for a closed chain. We will 
consider the case $0<V\le t'$, where we can use the parametrization 
$\cos \eta = V/t'$. It is interesting to note that for this case the 
Hamiltonian can be mapped with the help of the Jordan-Wigner transformation 
onto that of an ``easy-plane'' antiferromagnetic spin-1/2 chain with 
Dzyaloshinskii-Moriya interaction, with $V-\mu$ playing the role of an 
external magnetic field  \cite{DM}. 
%%%%%%%%%%%%%%%%%%%%%%%%%%%%%%%%%%%%%%%%%%d
\begin{figure}
\begin{center}
\vspace{-15pt}
\includegraphics[scale=0.4]{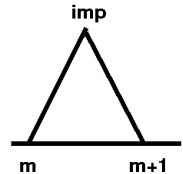}
\end{center}
\vspace{-15pt}
\caption{Illustration of the coupling of the impurity to the host. The impurity 
is coupled to its neighboring sites with a hopping amplitude 
$f(\theta, \eta)\cosh(\theta) \,t $ and interaction $f(\theta, \eta)V$, with 
$f(\theta, \eta)$ and $V$ defined in the text.}   
\label{imp}
\end{figure}
%%%%%%%%%%%%%%%%%%%%%%%%%%%%%%%%%%%%%%%%%%%

Let us now introduce an impurity by adding a lattice site, labeled $imp$ and 
located, say, between sites $m$ and $m+1$ of the chain. To maintain 
integrability of the theory, the coupling of the impurity site to the host has 
to be chosen judiciously. Using a template from Ref.~\onlinecite{KZ} and 
adapting it to the present case, we are led to the following form of the 
impurity Hamiltonian:  
\begin{eqnarray} \label{Himp}
{\cal H}_{\text{imp}}\!&\!=\!&\! f(\theta, \eta) \big(h_{m,imp} +h_{imp,m+1} 
- h_{m,m+1}
\nonumber \\
&& \ \ \ \ \ \ \ \ \  -g(\theta, \eta) [h_{m,imp},h_{imp,m+1}] \big) \ , 
\end{eqnarray} 
where $[... \, , ...]$ denotes a commutator, $f(\theta,\eta) \equiv 
\sin^2 \eta /[\sinh^2 \theta \!+\!\sin^2 \eta], g(\theta,\eta)\! \equiv \!
i \tanh \theta/ \sin \eta$, and where $h_{m,imp}$ and $h_{imp,m+1}$ have the 
same structure as in Eq.~(\ref{Hhom}) but with 
$t \rightarrow t_{\text{imp}}\equiv t \cosh \theta$. The real parameter $\theta$ 
defines the coupling of the impurity to the host. The case $\theta =0$ simply 
corresponds to the addition of a lattice site with no other modification, while 
for $\theta \to \infty$ the impurity site decouples from the host. Note that 
for any $\theta \neq 0$ the hopping and interaction between the neighboring 
sites $m$ and $m+1$ also get modified by ${\cal H}_{\text{imp}}$. It is worth 
pointing out that the structure of the impurity Hamiltonian becomes much 
simpler for the case of an open chain with the impurity situated at its edge: 
For that case we have ${\cal H}_{\text imp} = f(\theta, \eta) h_{M,imp}$ 
(where $M=L/a$ labels the last site in the chain, with $L$ the length of the 
chain and $a$ the lattice spacing). Also note that the commutator term in 
Eq.~(\ref{Himp}), while necessary for integrability, is irrelevant from the 
point of view of the renormalization group and can be neglected in the 
long-wavelength limit \cite{AndreiDestri1}. It can be checked that the gauge 
transformation, which removes the phase shift $2\pi \phi$ from the Hamiltonian 
for open boundary conditions and transfers it to twisted boundary conditions 
for the closed chain, can be applied also when the impurity interaction in 
(\ref{Himp}) is included. \\

\section{Periodic chain: Finite-size spectrum from Bethe Ansatz}
While the impurity Hamiltonian in Eq. (\ref{Himp}) breaks lattice translational invariance, 
single-particle backscattering (reflection) is not possible. This is a necessity
for the applicability of the BA method
to which we now turn. For the case of a periodic chain, with the SOI encoded 
by twisted boundary conditions, we obtain the BA equations (cf. the corresponding equations for 
the homogeneous chain without SOI \cite{SSSU})
\begin{eqnarray} 
&&e_{1}(\lambda_{\alpha}+ \theta)e_1^M(\lambda_{\alpha}) (-1)^{-{M\over 2} -N} 
e^{i2\pi\phi} \nonumber \\
&&= - \prod_{\beta=1,\beta \ne \alpha}^N e_2(\lambda_{\alpha}- \lambda_{\beta})
\label{BAE}
\end{eqnarray}
which determine the quantum numbers $\{\lambda_{\alpha}\}_{\alpha=1}^N$ (with 
$N$ the number of electrons) that parametrize the eigenfunctions and 
eigenvalues
\begin{eqnarray}
&&E=E_0 - \sum_{\alpha=1}^N \left(V-\mu -t'{\sin^2 \eta\over \cosh 
\lambda_{\alpha} -\cos \eta} \right) 
\label{BAEn}
\end{eqnarray}
of the total Hamiltonian ${\cal H} = {\cal H}_{\text{wire}} 
+ {\cal H}_{\text{imp}}$. Here $\alpha=1,\dots, N$, and $e_n(x) = 
\sinh [(x+in\eta)/2]/\sinh [(x-in\eta)/2]$, with $E_0 =MV/4$. In the 
noninteracting limit $\theta = V = 0$, the quantum numbers 
$\{\lambda_{\alpha}\}_{\alpha=1}^N$ become ordinary rapidities connected to the 
crystal momenta, and one recovers the expected result for noninteracting 
spinless fermions with an SOI. Less trivial is the property that the BA equations in
(\ref{BAEn}) are blind to the position of the impurity \cite{KondoComment}. This feature,
signaling that the impurity is non-reflecting, appears also in the related problem of ``mobile"
integrable impurities \cite{Tsukamoto,kriv}. As we shall see, it has dramatic consequences 
for correlation functions and observables.

The parameter $\theta$, determining the coupling of the impurity to the host, 
introduces a low-energy scale $T_{\theta} \sim t \exp(-\pi |\theta|)$, 
analogous to a Kondo temperature \cite{Kondo}. It defines a crossover between 
a low-energy regime where the impurity site is strongly coupled to the host, 
and a high-energy regime with the site being ``asymptotically free''. 
Importantly, the scale $T_{\theta}$ characterizes how the impurity influences 
the zero-frequency response of the system to an applied electric field: At low 
temperatures, $T \ll T_{\theta}$, one finds that the impurity contribution 
$\kappa_{\text{imp}}$ to the charge stiffness behaves as $\kappa_{\text{imp}} 
\sim 1/T_{\theta}$, while at high temperatures, $T \gg T_{\theta}$, 
$\kappa_{\text{imp}} \sim 1/T\cosh^2(\mu/2T)$, in both cases with corrections 
$\sim 1/\ln(T/T_{\theta})$ for $V=t$. It is important to point out that the 
appearance of the energy scale $T_{\theta}$ hinges on the presence of the 
interaction $\sim V$ in Eq. (\ref{Himp}). This is different from the  
archetypal Anderson single-impurity model in which the charge sector does not 
feature a crossover scale \cite{Kondo}.

The SOI shows up twofold in the BA equations (\ref{BAE}), as a renormalization
of the hopping $t$ due to the SOI amplitude $\alpha$ and in the phase factor 
$\exp(i2\pi\phi)$. Their influence on persistent currents and correlation functions
is most easily obtained via the finite-size corrections to the 
energy \cite{Zb}. The derivation of the finite-size corrections $\Delta E$ for the 
homogeneous model in Eq. (\ref{Hhom}) follows standard routes. To leading order
in $1/L$, 
\begin{equation}
\Delta E = {2\pi v\over L}\Delta\, ,  
\label{finEn}
\end{equation}
with $\Delta = [2Z]^{-2} (\Delta N)^2 + Z^2[D-\phi]^2 \, +\, n^+ + n^-$, and 
$v$ the velocity of low-lying excitations at the Fermi points. Here $Z$ is 
the ``dressed charge'' \cite{Zb}, connected to the ground state charge 
stiffness $\kappa(\mu)$ by $Z^2 = \pi v \kappa(\mu)$ and taking values from 
$\sqrt{\pi/2(\pi-\eta)}$ to 1 as $\mu$ increases from $V$ to $t'+V$ (where the 
number of electrons becomes zero). The quantum numbers, $\Delta N$, $D\, 
( = \!\Delta N/2 \, \mbox{mod} \, 1)$ and $n^{\pm}$ keep track of particle 
excitations, excitations from one Fermi point to the other (from umklapp), and 
particle-hole excitations, respectively. 

Let us now see how the result in Eq. (\ref{finEn}) gets modified when adding 
the impurity. An analysis similar to that for the homogeneous model yields the 
same expression for $\Delta E$ as in Eq. (\ref{finEn}), but with 
$\Delta \to \Delta_{\text{imp}}$, where 
\begin{equation}
\Delta_{\text{imp}} = [2Z]^{-2}[\Delta N - n_{\text{imp}}]^2 + Z^2
[D -\phi -d_{\text {imp}}]^2 \ ,    
\label{finEnsimp}
\end{equation}
where $n_{\text{imp}} = \int_{-\Lambda}^{\Lambda}d\lambda \rho(\lambda)$ is the 
valence of the impurity site and
\begin{equation}
d_{\text{imp}} = {1\over2} \left( \int_{-\infty}^{-\Lambda} d \lambda 
\rho(\lambda) -  \int^{\infty}_{\Lambda} d \lambda \rho (\lambda)
\right) \ .
\end{equation}
Here $\rho (\lambda) $ satisfies the integral equation
\begin{equation}
\rho(\lambda) %+ \rho_{1,h}^{(1)}(\lambda) 
= a_{1}(\lambda -\theta) 
-\int_{-\Lambda}^{\Lambda} d \lambda' a_2(\lambda -\lambda')
\rho (\lambda') \ , 
\label{rhoimp}
\end{equation}
where $a_{n}(x)\! \equiv \!2\partial_x (\tan^{-1} [\cot (n \eta/2) \tanh (x/2)])$,
and the integration limits $\pm \Lambda$ play the role of Fermi points. 
Note that the values of $n_{\text{imp}}$ and $d_{\text{imp}}$ are defined mod 1. \\

\section{Correlation functions for the periodic chain}
Given the results in Eqs. (\ref{finEn}) and (\ref{finEnsimp}), one can now 
calculate the persistent current \cite{pers} (Aharonov-Bohm-Casher effect 
\cite{ABC}) by differentiating the finite-size correction to the ground state 
energy with respect to the external flux (which can be introduced similar to 
$\phi$). Here, we instead focus on how to obtain asymptotics of correlation 
functions. The method for this is well-known, and uses conformal field theory
(CFT) \cite{YB} to take advantage of the conformal symmetry underlying the model.
Introducing the conformal dimensions 
$\Delta^{\pm}$, a correlation function for an operator ${\cal O}$ in the 
ground state of the closed homogeneous chain can be written as 
\begin{equation} 
\langle {\cal O}(x,t) {\cal O}(0,0)\rangle \sim 
{e^{-2i(D-\phi)k_Fx}\over (x-ivt)^{2\Delta^+}(x+ivt)^{2\Delta^-}} \ , 
\label{cor1}
\end{equation}
where $k_F = \pi N/2L$ is the Fermi wave number, and with the distance $x=ja$ satisfying $a \ll x \ll L$ with $j$ an integer. 
For small nonzero temperatures $T$ one has to replace 
$(x\mp ivt)$ by $v \sinh [\pi T (x\mp ivt)/v]/\pi T$ in Eq.~(\ref{cor1}). By {\em Cardy's formula} \cite{Cardy}, 
the conformal dimensions $\Delta^{\pm}$ are related to $\Delta$ in Eq.~(\ref{finEn}) by $\Delta = \Delta^+ + \Delta^-$
with $\phi$ absorbed in a twisted boundary condition on the operator ${\cal O}$ \cite{Tsukamoto}, as manifest in Eq. (\ref{cor1}).
One thus obtains for the homogeneous model without impurity
\begin{equation}
\Delta^{\pm} = \frac{1}{2}\biggl[ZD \pm {\Delta N\over 2Z}\biggr]^2  
+ n^{\pm} \ .
\label{confhf}
\end{equation}
For the density-density correlation function the choice of quantum numbers is 
$\Delta N = 0$ with $D$ a nonzero integer \cite{Zb}. It follows that the long-time dynamical density 
correlations are given by
\begin{equation} 
\langle n(x,t)n(x,0)\rangle = n_c +\mbox{const.}\times t^{-\gamma_1} + ...
\label{nn}
\end{equation}
where $\gamma_D \equiv 2(ZD)^2$ and $n_c$
is a constant. 
%The subleading correction, denoted ``...'', is obtained by invoking the 
%``descendants'' $n^+ = n^- = 1$ for $D=0$. 
For the single-electron Green's function we 
must instead choose $\Delta N = 1$ with $D$ half-odd-integer \cite{Zb}. We thus obtain,
\begin{equation} 
\langle c(x,t)c^{\dagger}(x,0)\rangle = \mbox{const.}\times t^{-\nu_{1/2,1}} + ...,
\label{dd}
\end{equation}
where $\nu_{D,\Delta N} = 2(ZD)^2 + (\Delta N)^2/2Z^2$.
As revealed by Eqs.~(\ref{nn}) and (\ref{dd}), the dependence of the dressed charge $Z$ on the renormalized coupling $t'=\sqrt{t^2 +\alpha^2}$ 
makes the SOI suppress large-time density-density correlations while the single-electron Green's function instead gets enhanced.

Adding the impurity, now considering the entire Hamiltonian ${\cal H} = 
{\cal H}_{\text{wire}} + {\cal H}_{\text{imp}}$, the theory is no longer invariant under
the full conformal group as the presence of the impurity breaks translational invariance. 
However, exploiting a {\em boundary CFT} approach \cite{AffleckPolonica}, we can still extract 
information about correlation functions using the following trick \cite{WongAffleck}: We fold the system in half at 
the impurity position $x\!=\!0$ (taken between sites $m$ and $m\!+\!1$ in Eq.\,(\ref{Himp})), and represent
left- (right-) moving electrons at $x\!<\!0$ by an auxiliary channel of electrons moving right $\!$(left) at $x\!>\!0$. 
Via this construction the impurity gets traded for a boundary condition at $x=0$ that is 
left invariant under a restricted set of conformal transformations and where a forward scattering process
(the only process allowed by ${\cal H}_{\text{imp}}$ in Eq. (\ref{Himp})) corresponds to having an electron 
come in through one channel and then reflected back through the other. 
As shown in the Appendix, the sum of the boundary scaling dimensions in the auxiliary problem for $x>0$ precisely defines the bulk scaling dimensions of the original problem, and one finds that these are identical to those of the homogeneous chain without the impurity. As a consequence, the long-time density and single-electron correlations in the presence of the impurity differ from those in Eqs. (\ref{nn}) and (\ref{dd}) only by the shift $\phi \rightarrow \phi + d_{\text{imp}}$.  This result signals the distinctive feature of an {\em integrable} quantum impurity embedded in a one-dimensional system: All particles impinging on the impurity are perfectly transmitted across the impurity site, with the scattering phase shift $d_{\text{imp}}$ absorbable into a twisted boundary condition.

Correlation effects become different when considering the dynamic response after a quantum quench at $t=0$, set off by suddenly switching on the impurity-electron interaction in Eq. (\ref{Himp}). As detailed in the Appendix, the impurity-renormalized boundary condition now implies that 
$\Delta N \rightarrow \Delta N - n_{\text{imp}}$ and $D \rightarrow D - d_{\text{imp}}$ in Eq. (\ref{confhf}). Thus, not only the amplitudes but also the exponents
$\gamma_D$ and $\nu_{D,\Delta N}$ get modified by the presence of the impurity. Remarkably, the large-time correlations are translationally invariant, insensitive to the particular location of the impurity. We conjecture that also this property reflects the integrability of the system: By the quantum quench, energy is transferred to the system via the impurity-electron interaction, and the perfectly transmitting impurity releases a finite-momentum excitation that runs around the ring and influences the correlations uniformly in the bulk. This picture is suggestive considering the structure of the BA equations, Eq. ~(\ref{BAE}), which makes it possible to associate nonzero momentum with the impurity. In this way, it effectively comes to play the role of a wave spreading over the ring, illustrating a kind of particle-wave duality. 

To elucidate the phenomenon, it may be 
useful to make an analogy with recent work on interacting 1D spinless fermions 
with nonlinear dispersion relations \cite{kriv}. Formally, the nonlinear 
corrections to the low-energy spectrum can be related here to the presence of 
a fictitious impurity with properties very similar to the one introduced in 
our model. In short, the difference between our impurity and the fictitious 
one is in the definition of the parameter $\theta$. For our case, $\theta$ is 
determined by the impurity-host coupling, whereas for the fictitious impurity 
$\theta$ is instead the rapidity of a high-energy excitation. With this 
observation, it also becomes easy to generalize our results for the correlation 
functions to include the nonzero curvature of the dispersion relation. We 
simply use the additivity of the $1/L$ corrections, and we add 
$n_{\text{imp}}^{f}(\Lambda_h)$ and $d_{\text{imp}}^f(\Lambda_h)$ (with $f$ 
denoting ``fictitious'') to $n_{\text{imp}}(\theta) $ and 
$d_{\text{imp}}(\theta)$, where $\Lambda_h$ defines the rapidity of the 
high-energy excitation. For  $\Lambda_h \sim \Lambda$,  
$n_{\text{imp}}^{f}(\Lambda)$ and $d_{\text{imp}}^f(\Lambda)$ (both determined 
mod 1) can be expressed in terms of the dressed charge $Z$ \cite{kriv}. In related,
earlier work, Tsukamoto {\em et al.} \cite{Tsukamoto} argued that the sudden insertion
of a mobile impurity into an interacting 1D electron system produces nontrivial bulk correlation
functions at large times when backscattering is suppressed, thus presaging our exact results via 
the analogy above. Interestingly, the new correlations produced by the quench are interpreted 
as being due to an orthogonality catastrophe \cite{Anderson} similar to that in the x-ray edge singularity 
for systems with a suddenly created localized core hole:  The screening effects due to the electrons lead 
to an ``infrared catastrophe", yielding a nontrivial asymptotic behavior of correlation functions in the 
long-time regime. \\

\section{Open chain}
Turning to the case of an open chain, with a local potential $h$ attached to 
its edges, the stratagem from above can be repeated step by step. We find the following for 
the finite-size corrections:
\begin{equation} 
\Delta E = {\pi v \over L}\Delta_b, 
\label{BS}
\end{equation}
with 
\begin{equation} 
\Delta_b = [2Z^2]^{-1}[\Delta N +\Theta(h,\theta)]^2 + n.  
\label{SD}
\end{equation}
By putting the impurity at one of the edges, choosing $m$ in Eq. (\ref{Himp}) as the
corresponding boundary site, $\Delta_b$ in Eq. (\ref{SD}) takes the role of boundary
scaling dimensions governing the large-time correlation functions in the neighborhood of the impurity. 
Here
\begin{equation}
\Theta(h,\theta) = -{1\over 2}\int_{- \Lambda}^{\Lambda}d\lambda \rho(\lambda), 
\end{equation}
with $\rho(\lambda)$ the solution of the integral equation
\begin{eqnarray}
\rho(\lambda) 
&=&  {1\over2}\biggl(\, \sum_{j=0,\pm 1} a_1(\lambda + j\theta) 
+ a_2(\lambda)\biggr) \\
&+&a_{\mu_h}(\lambda) \!- \!\int_{-\Lambda}^{\Lambda} \!d\lambda' a_{2}(\lambda\! 
-\!\lambda')\rho(\lambda'). \nonumber
\end{eqnarray}
The inhomogeneous term $a_{\mu_h}$ is determined as $a_n$ above with the 
formal substitution $n \rightarrow \mu_h$, with $\mu_h =   \ln [g_-(\eta,h/t)/g_+(\eta,h/t)]^{1/2}$ 
an {\em effective} boundary potential determined by $h$, and with 
$g_{\pm}(\eta,h/t) \equiv \sinh [\ln \sqrt{\cos \eta \pm (2h/t)} 
\pm i\eta/2]$. Since there is now only a single Fermi point, 
$D \rightarrow 0, n^{\pm} \rightarrow n$, as manifest in Eqs.~(\ref{BS}) and 
(\ref{SD}). The nonappearance of the phase $\phi$ reflects the trivial 
topology of the open chain, with the spin-conserving SOI only renormalizing 
the hopping amplitude $t$. Given our results for the bulk correlations
in the periodic chain, we conjecture that the boundary correlations governed by
$\Delta_b$ are insensitive to a displacement of the integrable impurity away from the boundary.
Unfortunately, a proof of this is not easily constructed within a boundary CFT formalism. \\

\section{Discussion}
In summary, using a combined Bethe ansatz and conformal field theory approach, we have obtained the exact asymptotic behavior of correlation functions in an integrable model of spin-polarized interacting electrons with a spin-conserving spin-orbit interaction. When the electrons are confined to a 
ring, the spin-orbit interaction tends to enhance the large-time
single-electron correlations while the density-density correlations get suppressed. After a sudden insertion of an integrable quantum impurity, with the impurity-electron 
interaction switched on abruptly, the scaling of the dynamic correlations picks up a nontrivial dependence on the presence of the impurity. The way the scaling dimensions depend on the spin-orbit coupling and the impurity phase shifts reveals that the quench enhances the long-time correlations, thus reducing the suppressing effect of the spin-orbit interaction on the density-density correlations. 
At large times, the phenomenon plays out with the same strength anywhere on the ring, independent of the distance to the impurity. We conjecture that this reflects the integrability of the impurity-electron interaction, which acts to produce a delocalized finite-momentum excitation after the quench, with electrons suffering only forward scattering off the impurity. Conceivably, the effect could be exploited in a future device for boosting electron correlations via a local quantum quench. The rapid progress in ``on-demand'' design of interactions in fermionic cold-atom systems holds promise for an experimental test.  
\\

\section*{ACKNOWLEDGMENTS}
We thank an anonymous referee for a pertinent question that helped us to significantly improve our manuscript. A.A.Z. acknowledges the support from the Institute for Chemistry of V.N.~Karasin Kharkov National University. H.J. was supported by the Swedish Research Council (Grant No. 621-2011-3942) and by STINT (Grant No. IG2011-2028). \\

\appendix
\section{BOUNDARY CFT FOR A PURELY TRANSMITTING IMPURITY}
In this appendix we show how to reformulate the problem of an integrable $-$ purely transmitting $-$ quantum impurity embedded in 
the bulk of a one-dimensional spinless fermion system so that Cardy's boundary conformal field theory (CFT) applies \cite{CardyReview}. 

Given the Hamiltonian in Eq. (1), we begin by taking a 
continuum limit, representing the lattice fermion operators $c_n$ by
\begin{equation} 
c_n \rightarrow \sqrt{a}[e^{ik_Fx}\psi_R(x) + e^{-ik_Fx}\psi_L(x)], \ \ x = na,  
\end{equation}
where $\psi_L$ and $\psi_R$ are chiral fields defined in the neighborhood of the Fermi points $k_F$ and $-k_F$, respectively, satisfying 
\begin{equation}  
\{\psi_{\lambda}(x), \psi^{\dagger}_{\lambda^{\prime}}(y)\} =  \delta_{\lambda,\lambda^{\prime}}\delta(x-y), \ \ \ \lambda, \lambda^{\prime} = L,R.
\end{equation}
Linearizing the spectrum around the Fermi points, the continuum limit of  Eq. (1) can then be expressed in current algebra form,
\begin{equation} \label{hamiltonian1}
H \!=\! \frac{v}{2} \int dx \left[\sum_{\alpha=L,R}\!:\!J_{\alpha}(x) J_{\alpha}(x)\!: + gJ_L(x) J_R(x)\right]
\end{equation}
with U(1) currents 
\begin{equation}
J_{\lambda}(x) = :\!\psi^{\dagger}_{\lambda}(x) \psi_{\lambda}(x)\!:,
\end{equation}
and where $v$ and $g$ are parameterized by $v_F$ and $V$ \cite{JJ}. The normal ordering : ... :  is carried out with respect to the filled Dirac sea. The Hamiltonian
in Eq. (\ref{hamiltonian1}) mixes left and right currents but can be diagonalized by the Bogoliubov transformation
\begin{equation}  \label{Bogoliubov}
J_{L/R} = \cosh \theta j_{L/R}(x) - \sinh \theta j_{R/L} (x)
\end{equation}
with $2\theta = \mbox{arctanh}(g/(v_F+g))$. One thus obtains 
\begin{equation} \label{hamiltonian2}
H = \frac{v}{2} \int dx \left[:\!j_L(x) j_L(x)\!: + :\!j_R(x) j_R(x)\!:\right],
\end{equation}
with the new currents satisfying the U(1) Kac-Moody algebra,
\begin{equation}
[j_{L/R}(x), j_{L/R}(y) ] = \pm i \delta^{\prime}(x-y).
\end{equation}
We now boost the currents into the complex plane $\{z = \tau + ix\}$ (with $\tau$ a Euclidean time) and identify the impurity site in Eq. (2) 
with the time axis $x=0$. Whereas the impurity-electron interaction in Eq. (2) is not easily expressible in terms of the currents, the current algebra formulation is still helpful for understanding how this interaction can be handled within the boundary CFT formalism. In this approach $-$ first used for a quantum impurity problem in Ref. \onlinecite{Affleck} $-$ the interaction in Eq. (2)  is traded for a conformally invariant boundary condition at $x=0$ \cite{AffleckPolonica}. As follows from the integrability of the model, in the present case the impurity is perfectly transmitting. This simplifies the problem. However, there is a catch: In the boundary CFT formalism no momentum or charge is allowed to pass through the boundary. To be able to use boundary CFT, we therefore have to reformulate the problem in such a way that our perfectly transmitting impurity gets represented by a perfectly reflecting boundary. The "trick" how to do this involves the introduction of an auxiliary channel of fermions, where pure transmission through the impurity site gets represented by pure reflection from one channel into the other \cite{WongAffleck}. Upon analytic continuation, one is left with two channels of left-moving (or right-moving) currents, both respecting translational invariance. The imprint of the impurity (which has now superficially disappeared from the problem) is seen in the new spectrum of scaling dimensions. These dimensions can be read off from the exact finite-size Bethe ansatz spectrum, thus providing access to the asymptotic correlation functions. 

To see how this blueprint plays out in mathematical terms, we first impose periodic boundary conditions on the transformed currents,
\begin{equation}  \label{periodic}
j_{\lambda}(\tau,0_+) = j_{\lambda}(\tau,0_-),
\end{equation}
thinking of the time axis $x=0$ as a boundary with periodic boundary conditions when there is no impurity present.
Next, we restrict the system to the interval $-L \le x \le L$ (taking $L \rightarrow \infty$ at the end), fold it in half, double the currents, and identify $x=-L$ and $x=L$. The new currents, defined in the semi-infinite complex plane $x \ge 0$, are connected to the old ones by
\begin{eqnarray} \label{DoubleCurrents}
j_{1L}(x) &\equiv &j_L(x),  \ \ \ \ \ j_{1R}(x) \equiv j_R(x) \\
\ \ \  j_{2L}(x) &\equiv& j_R(-x), \ \ j_{2R}(x) \equiv j_L(-x)
\end{eqnarray}
where we have suppressed the common time argument. As a consequence, the periodic boundary condition in Eq. (\ref{periodic}) takes the form  
\begin{equation}  \label{newperiodic}
j_{1L}(0) = j_{2R}(0), \ \ j_{2L}(0) = j_{1R}(0).
\end{equation}
By this procedure, the Hamiltonian in Eq. (\ref{hamiltonian2}) is now defined for $x\ge0$ only. Using Eq. (\ref{DoubleCurrents}) and the boundary condition in Eq. (\ref{newperiodic}), however, we can analytically continue the left-moving currents to $x<0$, with
\begin{equation}  \label{identification}
j_{1L}(-x) = j_{2R}(x), \ \ j_{2L}(-x) = j_{1R}(x),
\end{equation}
and then write the Hamiltonian in the full complex plane in terms of left-moving currents only (after having taken $L \rightarrow \infty$):
\begin{equation} \label{Chiral}
H = \frac{v}{2} \sum_{i=1,2} \int dx :\!j_L(x)j_L(x)\!: .
\end{equation}
We now bring in the impurity-electron interaction, Eq. (2). Introducing the notation $\{\Delta_L\}$ for the subset of chiral (``left-moving", say) scaling dimensions that make up the {\em boundary scaling dimensions} for a given boundary condition, Cardy's finite-size boundary formula is expressed as  $E = E_0 + \pi v \Delta_L/\ell$ \cite{Cardy}. Adapting it to our case with two copies of left-moving channels, indexed by $i=1,2$, we have
\begin{equation} \label{Cardy}
E = E_0 + \frac{\pi v}{\ell}( \Delta_1 + \Delta_2).
\end{equation}
This formula connects the energy spectrum of the theory on the strip $\{w = u+iv\}, 0 \le v \le \ell, -\infty< u< \infty,$ to the sum of boundary scaling dimensions $\Delta_1 + \Delta_2 = 2\Delta_L$ in the semi-infinite plane $\{z = \exp(\pi w/\ell)\}$ associated with the
boundary condition at $x=0$ which emulates the impurity interaction \cite{footnote1}.  It is crucial here to realize that the images of this boundary condition at the two edges of the strip effectively correspond to the insertion of {\em two} copies of the impurity. While in our case we are not able to pinpoint the appropriate boundary condition {\em per se},
having obtained the exact finite-size spectrum from the Bethe ansatz, solution we can nonetheless identify the spectrum of scaling dimensions using Eq. (\ref{Cardy}): The subset of chiral scaling dimensions $\{\Delta_L\}$ that corresponds to the new boundary condition is simply selected via inspection of the finite-size Bethe ansatz spectrum after insertion of two auxiliary impurities in each channel, one at each edge of the strip. It follows that the quantum numbers $\Delta N$ and $D$ in Eq. (10) 
get renormalized {\em twice}, with $\Delta N \rightarrow \Delta N - n_{\text{imp}}(v\!=\!0) = \Delta N^{\prime}$ 
and $D \rightarrow D - d_{\text{imp}}(v\!=\!0) =  D^{\prime}$ from  the $v=0$ edge, and $\Delta N^{\prime} \rightarrow 
\Delta N^{\prime} + n_{\text{imp}}(v\!=\!\ell) = \Delta N$ and $D^{\prime} \rightarrow D^{\prime} + d_{\text{imp}}(v\!=\!\ell)= D$ 
from the $x = \ell$ edge. The opposite signs of the charge valences $n_{\text{imp}}$ and level shifts $d_{\text{imp}}$ 
at the two edges here originate from the opposite signs of the phase shifts at 
$v = 0$ and $v= \ell$ (corresponding to $\tau<0$ and $\tau>0$ respectively in the semi-infinite plane). In contrast, when the impurity interacts with the fermions only when $\tau\ge0$, as after a quantum quench at $\tau=0$, only the boundary condition at the corresponding edge of the strip, $v=0$, gets renormalized. As a result, the dynamic correlation exponents pick up a nontrivial contribution from the impurity, with $\Delta N \rightarrow \Delta N - n_{\text{imp}}$  and $D \rightarrow D - d_{\text{imp}}$. As was made explicit in our analysis above, the second channel of left-moving currents in Eq. (\ref{Chiral}) simulates the right-moving currents in (\ref{hamiltonian2}). Therefore, {\em bulk scaling dimensions} $\{\Delta\}$ appear in Eq. (\ref{Cardy}), disguised as sums of chiral scaling dimensions labeled by the channel index: $\Delta = \Delta_1 + \Delta_2$.
It is important to emphasize that this conclusion is certain to be valid only for an {\em integrable} impurity, since only for this case are we ensured that the impurity is purely transmitting in the basis of the $j_{L/R}(x)$ currents that diagonalizes the bulk interactions, thus maintaining the decoupling of the two channels. 

As a concluding remark in this appendix, it is important to realize that it is precisely the absence of backscattering from the integrable impurity that causes {\em all} large-time dynamic correlation functions to be governed by the {\em same} scaling dimensions $-$ independent of the distance from the impurity. Hence there is no crossover from bulk to boundary critical behavior as one approaches the impurity site. The breaking of translational invariance due to the impurity shows up only in a shift of the phase of the full space-time correlation function in Eq. (\ref{cor1}). Clearly, as emphasized throughout this work, this feature is not generic but crucially hinges upon the design of the impurity interaction, having made it integrable and therefore purely transmitting.

\end{document}